\documentclass[aip,rsi,reprint]{revtex4-1}
\usepackage{graphicx}
\draft

\begin{document}

\title{Ultra-low supply voltage crystal quartz oscillator}
\author{A.M. Korolev}
\email{korol.rian@gmail.com.}
\affiliation{Institute of Radio Astronomy, NAS of Ukraine, Chervonopraporna St. 4, Kharkiv 61002, Ukraine.} %
\author{V.M. Shulga}
\email{shulga@rian.kharkov.ua.}
\affiliation{Institute of Radio Astronomy, NAS of Ukraine, Chervonopraporna St. 4, Kharkiv 61002, Ukraine.} %
\affiliation{International Center of Future Science,
Jilin University, Qianjin Street No. 2699, Changchun, Jilin Province, 130012, China} %
\author{O.G. Turutanov}
\email{turutanov@ilt.kharkov.ua.}
\affiliation{B. Verkin Institute for Low Temperature Physics and Engineering, NAS of Ukraine, Nauky Ave. 47, Kharkiv 61103, Ukraine.} %

\date{\today}

\begin{abstract}
An ultra-low-voltage crystal quartz oscillator is proposed. The
approach to its design is essentially based on using a HEMT
operating in unsaturated dc regime and a quartz resonator as a
resonant impedance transformer. The 25 MHz prototype shows steady
oscillations at the supply voltage of less than 17 mV and the power
consumption as low as 300 nW, i.e., 1-2 orders of magnitude lower
than other to-date oscillators. This approach is good for building
ultra-low consumption radio devices including those working at low
temperatures.
\end{abstract}


\maketitle


Reduction of the supply voltage is one of the most distinct modern
trends in the development of digital and analog low-power
electronics, both in mobile and stationary applications. New
semiconductor technologies and circuitries are currently under
development \cite{1Matheoud,2Sarkar}. Quartz crystal oscillator is a
core functional unit which is of strong interest to developers from
the point of view of lowering its consumption power and supply
voltage \cite{3Priasmoro,4Badal}. To date, the supply voltage as low
as 0.2 V and consumption power down to 7 $\mu$W for the radio
frequency quartz crystal oscillator are achieved \cite{3Priasmoro}.
In our paper, we propose a specific and effective approach to design
of an ultra-low-supply-voltage quartz crystal oscillator. The main
goal of the work is to determine the limitation in the supply
voltage reduction.

Considering a generator as an amplifier with positive feedback, one
should choose an active element that has maximal gain at a minimal
supply voltage. We believe a field-effect transistor (FET) to be the
most promising candidate for such an element. The FET's current
channel contains no potential barrier (p-n junction). Accordingly,
the current-driving drain-source voltage $U_{ds}$ has no threshold.
But more important thing is the electrostatic principle of the
channel conductivity control. At frequencies much lower than the
FET's cut-off frequency $F_t$ the transistor control circuit
(gate-source) consumes very low active power that ensures high
current gain. This provides noticeable power gain $G_p$ even at the
voltage gain $G_u$ being much smaller than the unity that is typical
for $U_{ds} \textless$ 0.5 V.

In papers \cite{5Korolev,6Korolev,7Korolev} devoted to FET operation
in the unsaturated mode, it was shown that the minimal required
drain-source voltage $U_{min}$ for a fixed $G_p$ decreases when
decreasing the operating frequency $F$ (more precisely, the $F/F_t$
ratio). When the FET's complex input impedance perfectly matches
that of a signal source, the $U_{min}$ is proportional to $F$
\cite{5Korolev,6Korolev}. Near-perfect matching of high-frequency
small-signal FETs is realizable at frequencies above, roughly, 100
MHz. The quasi-matching is more common for lower radio frequencies
when the magnitudes of the source and source-gate FET's impedances
are equal. In this case \cite{7Korolev}, $U_{min}\sim \sqrt{F}$.
These laws are principally due to the electrostatic modulation of
the conduction of any-FET current channel while the exact frequency
variates depending on specific class of FETs. The net parameter to
compare the amplifying characteristics of differently designed FETs
is the cut-off frequency $F_t$ determined by the gate capacitance
$C$ and the transconductance $G_m$ of the FET: $F_t=G_m/2 \pi C$.
The boundary frequencies for the most modern heterostructural FETs
(HEMTs) are hundreds of gigahertz, so there is an undoubted
possibility of a dramatic decrease in $U_{min}$ for frequencies
ranging from tens to hundreds of megahertz. As to amplifying
devices, the feasibility of ultra-low voltage and power supply has
been demonstrated repeatedly
\cite{5Korolev,6Korolev,7Korolev,8Oukhanski,9Noudeviwa}. However, in
the vast majority of cases, they were ultra low-noise cryogenically
cooled amplifiers and oscillators \cite{2Sarkar}. Here we discuss
ultra-low power (ULP) self-excited oscillators that are operable
both at cryogenic temperatures and under normal temperature
conditions. The first results will be presented here to confirm the
possibility of creating an important kind of electronic units,
quartz oscillators with ultra-low supply voltage (10 to 100 mV) and
consumption power (a few microwatts or less) for promising
ultra-efficient radio devices.

The basis of our approach is to use a HEMT in a dc regime close to
the unsaturated one. This mode, in contrast to the saturated
microcurrent regime widely used in micropower electronics, is
associated with moderate load values (hundreds of ohms) thus making
it possible to build high-frequency devices. The near-unsaturated dc
regime assumes the transistor operating point to be located in the
upstream region of the current-voltage characteristics family. For
this area, the transistor self voltage gain ($G_u$, the product of
the channel differential resistance by the transconductance) does
not exceed unity in the linear amplification mode. When the
transistor works as an active element of an oscillator, its
operating point moves along so-called ``limit cycle'' and can enter
both drain current cut-off and fully open channel areas. The
cycle-averaged $G_u$ will further decrease. Therefore, the current
amplification and the feedback loop which provides the necessary
output-input impedance matching with consequently greater-than-unity
loop gain are required. Let us consider the impedance matching task
in more detail.

The perfect match in our case means maximizing $G_p$. Hence, the
impedances of the transistor gate-source control circuit and the
connected "signal source" circuit should be complex-conjugated. In
first approximation, the input circuit of a low-power HEMT with $F_t
\approx$ 50 GHz is modeled by a capacitor $C \approx$ 0.5 pF and a
resistor $R_{gs} \approx$ 5 Ohm connected in series. It is clear
that the matched source ($LR$ circuit) must have quality factor of
more than 1000 ($F$ = 30 MHz) that is not physically feasible with
lumped-parameter elements. Even inductor with reactance of the order
of 10 k$\Omega$ is unrealizable since its own resonance would lie in
the frequency range below the operating one. The option of a
distributed device (a resonator with a length of about 1 meter) is
obviously nonsense.

It is the crystal quartz resonator that actually has the required
parameters. As is well known, its equivalent circuit is a
high-quality oscillatory tank circuit. By connecting such an element
to the input (gate-source) of the transistor and partly coupling it
to the output (drain-source), we implement a matching resonant
impedance transformer. The phase condition for self-excitation is
fulfilled by introducing a capacitive feedback element between the
drain and the source of the transistor, similarly to the Colpitts
oscillator.

The impedance matching performed in this way does not provide the
perfect complex conjugation. It is the quasi-matching when only
impedance magnitudes are equal which is worse for reducing the
minimum power supply voltage ($U_{min}\sim F^{0.5}$, see above).
Below we give a description of a possible circuitry implementation
of the crystal oscillator based on all these considerations.

\begin{figure}[h!]
\centering
\includegraphics{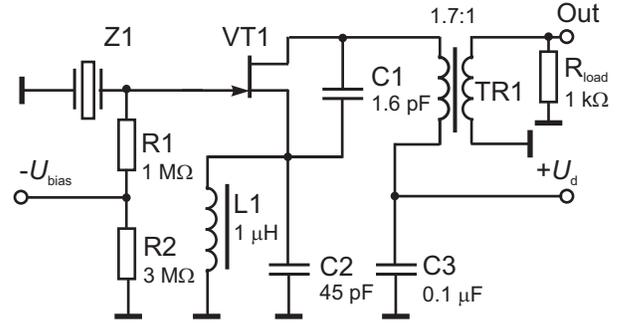}
\caption{\label{fig01}%
Circuit diagram of the oscillator. VT1 is AVAGO ATF36077.  Z1 is a
general-purpose 25 MHz quartz resonator. The connection between Z1,
R1 and VT1 gate should be made as short as possible in the form of
"air bridge" not touching the PCB substrate.}
\end{figure}

A schematic diagram of the oscillator is shown in Fig.~\ref{fig01}.

The circuit provides generation at the fundamental resonance of the
crystal from 15 to 30 MHz.

The preliminary circuit calculation was performed using the
datasheet from the transistor manufacturer and additionally measured
static characteristics in the unsaturated region.

The bias circuit R1, R2 (dissipation power of about 0.1 $\mu$W) is
used to adjust the optimal dc regime (gate-source voltage
$U_{gs}$=-0.4...-0.5 V at room temperature and -0.2...-0.3 at 78 K).
Also, for practical applications, it is possible to select a
zero-bias transistor for VT1.

As mentioned above, the objective of this study is to find the lower
limit for reduction of the supply voltage and, correspondingly, the
power consumption.

\begin{figure}[h!]
\centering
\includegraphics[width = 1.0\columnwidth]{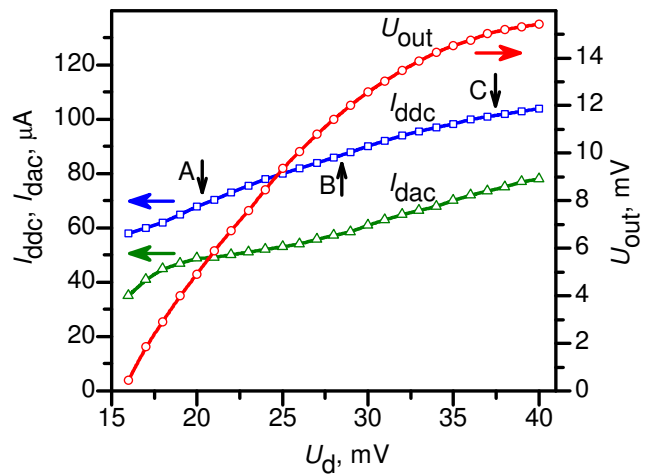}
\caption{\label{fig02}%
Static $I_{ddc}(U_{ds})$ and dynamic $I_{dac}(U_{ds})$
characteristics of the transistor and output peak-to-peak voltage
$U_{out}$ of the oscillator vs. supply voltage $U_{d}$,
$U_{ds}=U_{d}$. For points A, B, C indicated by vertical arrows,
transconductance $G_m$ and voltage gain $G_u$ are 0.5 mS and 0.15,
0.7 mS and 0.35, 0.9 mS and 0.6, correspondingly.}
\end{figure}

Fig.~\ref{fig02} displays 1) the dc drain current $I_{ddc}$ of the
quenched oscillator, 2) the dc drain current $I_{dac}$ of the
self-excited oscillator vs. the drain-source voltage $U_{ds}$, and
3) the generated signal amplitude $U_{out}$ vs. the supply voltage
$U_d$. Actually, $U_{ds}=U_d$. It is easy to see from $U_{out}$
curve that the stable generation is observed at $U_d=$17 mV that
corresponds to sub-microwatt (300 nW) power consumption. The
transconductance $G_m$ and the voltage gain $G_{u} =G_{m} (dI_{ddc}/
dU_{d}) $ are shown for some $U_d$ values indicated by letters A, B,
C. Such characteristics were not achieved earlier in the devices
with similar functionality.

Notably, the self voltage gain (with no load taken into account) for
the entire I-V characteristic does not exceed 1 being close to 0.1
in the vicinity of the threshold region. This indicates the
effective operation of the quartz resonator as an impedance
transformer which enables the balance of amplitudes and the
self-excitation. Note that, by indirect evidence, there is a very
high-quality loaded resonance. Indeed, reducing the resistance of R1
from 1 M$\Omega$ down to 0.5 M$\Omega$ causes increase in the
generation threshold by 5 mV.

The output signal waveform is close to sine because of filtering by
the output transformer. The contribution of harmonics does not
exceed -8 dB at supply voltages in the range of 15...50 mV. The
signal amplitude is about a 0.3 of the supply voltage at a 1
K$\Omega$ load. Power dc to rf conversion efficiency varies from 1\%
at $U_d$=20 mV to 4.5\% at $U_d$=33mV.

Because of high level of $1/f$ noise attributed to HEMTs, we found
it reasonable to make some spectral noise density measurements. For
a rough estimate of the spectrum, here are some quantitative
characteristics. Normalized oscillator spectral noise density S,
dBn/Hz is -71, -93, -108, -113 dB at detuning 10, 100,1000, and
10,000 Hz, respectively. The accuracy of a measurement is about 2
dB. The observed level of the phase noise at small detuning really
exceeds significantly that of specialized devices based on silicon
transistors and specially manufactured high-Q quartz resonators
\cite{10Apte}. Nevertheless, the level of the phase noise itself is
quite acceptable for general-purpose generators. Note that the
general concept of our device is by no means focused solely on the
use of HEMT. Silicon FETs with a high boundary frequency are also
applicable. The HEMT is particularly interesting because it can
operate at arbitrary low temperatures. So, the HEMT-based oscillator
can be a part of complex cryoelectronic devices, especially taking
into account its ultra-low power consumption since refrigerating
capacity of cryocoolers is very small, just microwatts below 0.1 K.
To test the presented device as an element of cryoelectronics, we
cooled it down to 78 K, with a corresponding correction of the bias
voltage (note that there is no significant change in the
amplification properties of the HEMT at lower temperatures). A
decrease, down to 10 mV, in the self-excitation threshold voltage
was observed, with power consumption of about 100 nW. Thus, the
oscillator can also be used in deep-cooled cryoelectronic devices,
especially taking into account that the brightness noise (electron)
temperature drastically falls down at the HEMT supply voltage lower
than 30 mV \cite{11Korolev}.

To summarize, the paper presents an approach to designing the
ultra-low-voltage crystal quartz oscillators. The concept is based
on the use of a field effect transistor in its unsaturated mode. To
implement the idea, a quartz resonator is exploited as a resonant
impedance transformer.

The steady self-excitation mode was observed with the supply voltage
of less than 17 mV and the power consumption as low as 300 nW, i.e.,
1-2 orders of magnitude lower than the modern records. The proposed
oscillator design can be recommended for use in ULP radio devices
(medical and other sensors, converters, RFIDs, etc.) including
cryoelectronic ones.

This work was supported by National Academy of Sciences of Ukraine
[grant number 0117U002398] and NATO SPS Programme through grant No.
G5796.

\par\medskip\textbf{DATA AVAILABILITY}\par\medskip

The  data  that  support  the  findings  of  this  study  are
available from the corresponding author upon reasonable request.

\end{document}